\documentclass[%
twocolumn,
 amsmath,amssymb,
aps,
prb,
]{revtex4-1}

\usepackage{graphicx,epsfig,epstopdf}
\usepackage{amsmath}
\usepackage{color}
\usepackage{caption}
\usepackage{subcaption}

\def\e{\begin{equation}}
\def\f{\end{equation}}
\def\_#1{{\bf #1}}
\def\.{\cdot}

\begin{document}

\title{Geometry free materials enabled by transformation optics for enhancing the intensity of electromagnetic waves in an arbitrary domain} 

\author{Ali Abdolali}
\email{Abdolali@iust.ac.ir}
\author{Hooman Barati sedeh}
\author{Mohammad Hosein Fakheri}

\affiliation{
	Applied Electromagnetic Laboratory, School of Electrical Engineering, Iran University of Science and Technology, Tehran, 1684613114, Iran}

\begin{abstract}
The concentration of electromagnetic waves (EM) is of utmost importance in many engineering applications such as solar-cells. According to the transformation optics (TO) methodology, a feasible approach for obtaining arbitrary shape concentrators is proposed. In contrary to the previous works, the obtained materials are homogeneous and independent of the concentrator shape. That is regardless of the input geometry always one constant material, which is optic-null medium (ONM), is used and the performance of device will not alter. This competency will ease the design process and circumvent the sophisticated calculations of the necessitating materials. To authenticate the concept, several numerical full-wave simulations were performed for different shape concentrators. It was observed that numerical results exhibit strong agreement with the theoretical investigations and corroborate the generality and effectiveness of the proposed designing method.

\end{abstract}

\maketitle

\section{Introduction}
TO methodology, which was proposed by Pendry \textit{et al.} \cite{pendry2006controlling}, has shown a great potential to design various devices that were deemed impossible to be achieved with conventional methods, such as invisibility cloaks \cite{schurig2006metamaterial,fakheri2017carpet}, field rotators \cite{kwon2008polarization}, collimating lenses \cite{jiang2011experimental,jiang2012broadband,wu2013transformation} and beam steering applications \cite{barati2018experimental}. Among these functionalities, the phenomenon of near-field concentration of EM waves plays a crucial role in the harnessing of energy in solar cells \cite{catchpole2008plasmonic}. Recently, several TO-based devices have been reported for this aim \cite{yang2010electromagnetic,rahm2008design,yang2009metamaterial,zhang2011cylindrical,zhao2018feasible,boardman2011nonlinear,li2015inverse,zhou2018perfect,sadeghi2015optical}. However, to the authors\textsc{\char13}    best knowledge, most of these works were limited to the circular-cross section concentrators which limit their practical usage for more general cases where there is a prerequisite to focus energy in an arbitrary region of interest. Although Yang \textit{et al.} proposed a method based on Laplace equation to achieve arbitrary shape concentrators , the propounded approach cannot be utilized in realistic situations since its obtained materials are inhomogenous and anisotropic \cite{yang2010electromagnetic}. However, Madni \textit{et al.} proposed a novel method based on multi-folded transformation optics that is competent to obviate the inhomogenity problem \cite{madni2018novel}. \par
In addition to the realization procedure, previously proposed mapping functions are strictly dependent on the desired concentrator shapes. That is if the  structure shape is changed, one must recalculate the necessitating materials and as a result the realization procedure should be repeated, which make these works impractical for real-life scenarios where reconfigurability is of utmost importance. All these challenges cause this  question to be asked that whether is there any alternative way of designing arbitrary shape concentrators that their necessitating materials are independent of the geometry? \par
Recently, Engheta \textit{et al.} proposed the concept of nihility media, as an environment where its constitutive parameters are near zero, to achieve various novel devices \cite{liberal2017near,edwards2008experimental}. After the introduction of near zero media, the presented concept was further became more generalized by defining it as a material obtained from transforming a volumless geometrical element such as point, line or surface to a finite volume of space \cite{yan2010generalized}. Among the materials obtained from these transformations, the one that is attained from mapping a surface into a volume is known as the optic-null medium (ONM) and attract scientists' attentions due to its unique characteristics. 
For instance, Sun \textit{et al.} utilized ONM for designing different functionalities such as waveguide bends \cite{sun2018waveguide}, wave front reshaping devices \cite{guo2016optical} and other interesting novel applications\cite{sun2018subwavelength,sun2017transformation,sun2016anti,sun2016optic,sun2016creating,sun2015optical}. Furthermore, it is well known that single conductor waveguides (e.g. rectangular waveguide) do not support transverse electromagnetic (TEM) mode\cite{balanis1999advanced}. However,  Catrysse \textit{et al.}  attained this mode from a waveguide which was filled with such a material \cite{catrysse2011transverse,catrysse2013routing}. The ONM was also used to achieve reflectionless diffraction-free bending and splitting devices by Zhang \textit{et al} \cite{zhang2017bending}.\par
In this paper, a new method to enhance the power intensity of EM waves in an arbitrary region of interest that is more feasible to be implemented in real-life scenario is proposed. Due to the propounded technique, the acquired materials become ONM, which obviates the demand of recalculating the necessitating materials when the shape of structure is changed.This will give the designer a more degree of freedom to enhance the EM waves power in any arbitrary region. First, the theoretical investigations of the method is given in details, then several numerical simulations are performed to demonstrate the capability of propounded approach in enhancing the EM intensity. We believe that not only this method could find applications in scenarios where high field intensity in an arbitrary cross section is needed, but also it is more feasible to be fabricated in comparison with its conventional counterparts, which were inhomogeneous and anisotropic.
\vspace{-0.5 cm}
\section{Theory}
In order to focus the incident wave in an arbitrary region of interest, the schematic diagram of Fig.1 is used as the space transformation,which three cylinders with arbitrary cross sections of $R_1(\phi)= \tau_1 R(\phi)$, $R_2(\phi)=\tau_2 R(\phi)$ and $R_3(\phi)=\tau_3 R(\phi) $ divide the space into three different regions. It should be noted that  $\tau_1$, $\tau_2$ and $\tau_3$ are constant coefficients which satisfy the condition of $\tau_1 < \tau_2 < \tau_3$  and $R(\phi)$ is an arbitrary continuous function with period of $2\pi$  that is specified with Fourier series as
\begin{equation}
R(\phi) = a_{0}+ \sum_{n=1}^{\infty} \{ a_{n} \cdot cos (n\phi) +b_{n} \cdot cos (n\phi) \}
\end{equation}
where $a_n$ and $b_n$ are constant coefficients that specify the contour shape.
\begin{figure}[!hb]
	\centering
	\includegraphics[width=\linewidth]{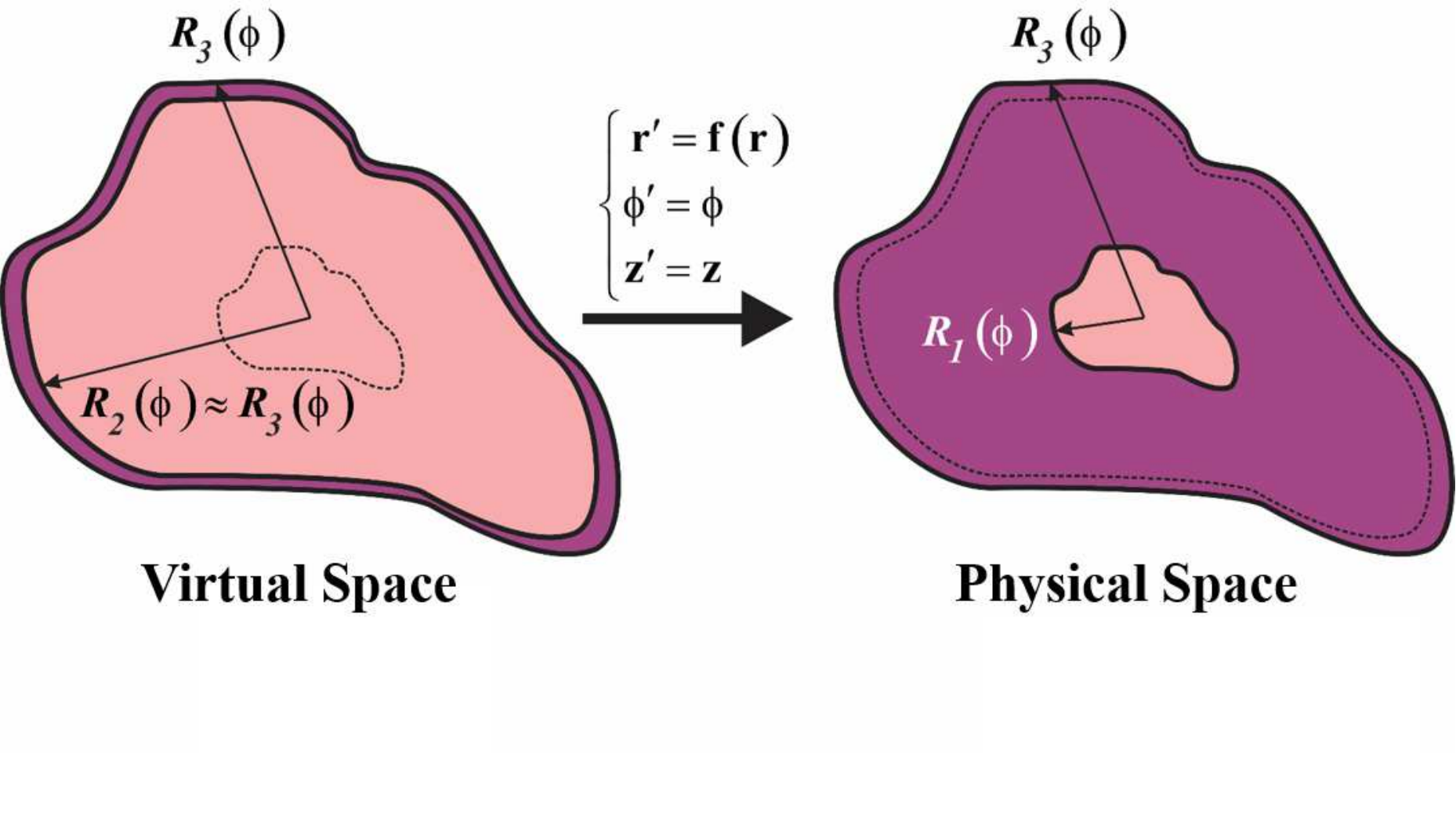}
	\vspace{-0.5 cm}
	\caption{The schematic of coordinate transformation for achieving arbitrary shape concentrators. }
	\label{fig:FIG_1}
\end{figure}

To concentrate the incident wave in a predefined region of $R_1(\phi)$, one must collect the energy that was originally located in $r<R_2(\phi)$ into the region of $r^{\prime} < R_1 (\phi^\prime)$. To this aim, the region of $r\in[0, R_2 (\phi)]$ must be compressed into the region $r^\prime\in[0, R_1 (\phi^\prime)]$, while at the same time $r\in[R_2 (\phi), R_3 (\phi)]$ is stretched into the region of $r^\prime\in[R_1 (\phi^\prime), R_3 (\phi^\prime)]$.
Since these two steps occur simultaneously, all the energy, which previously located in $r < R_2 (\phi)$ is now localized in the region of $r^{\prime} < R_1 (\phi^\prime)$  and as a result, the power intensity will be increased in the mentioned domain. It is notable to mention that as the azimuthal direction is not transformed in this kind of mapping (i.e., $\phi =\phi^\prime$); thus, the function which is competent to perform such a transformation could be expressed as
\begin{equation}
\   \left\{
\begin{array}{ll}
f_c(r,\phi)= \frac{\tau_1}{\tau_2}r& r'\in [0, R_1(\phi))\vspace{0.5 cm} \\
f_s(r,\phi)= \Omega \times r + \Psi \times R(\phi)  & r'\in[R_1 (\phi), R_3 (\phi)]
\end{array} 
\right. \
\end{equation}
where subscripts of $c$ and $s$ represent the compressed and stretched region, respectively and $\Omega=(\tau_3-\tau_1)/(\tau_3-\tau_2)$, $\Psi= [(\tau_1-\tau_2)/(\tau_3-\tau_2)]\tau_3$. According to the optical transformation theory \cite{pendry2006controlling}, the constitutive parameters of the physical space could be expressed as 
\begin{equation}
\frac{\varepsilon^{\prime}}{\varepsilon_0}=\frac{\mu^{\prime}}{\mu_0}= \frac{\Lambda \times \Lambda^T}{det(\Lambda)} 
\end{equation}
where $\varepsilon_0$ and $\mu_0$ are the constitutive parameters of virtual space which is assumed to be air and $\Lambda=\partial(x^\prime,y^\prime,z^\prime)/\partial(x,y,z)$ is the Jacobean matrix which relates the metrics of the physical space, $(x^\prime,y^\prime, z^\prime)$ to the virtual space, $(x,y,z)$. By substituting Eq. (2) into Eq. (3) the necessitating materials for each region will be achieved as
\begin{subequations}
	\begin{equation}
	\label{eq-a}
	\frac{\varepsilon'_{c}}{\varepsilon_0}=\frac{\mu'_{c}}{\mu_0}=
	\begin{bmatrix}
	1 &
	0&
	0 \\
	0 &
	
	1 &
	0\\
	0 &
	0 &
	(\tau_2/\tau_1)^2
	\end{bmatrix} 
	\end{equation}
	\begin{equation}
	\label{eq-b}
	\frac{\varepsilon'_{s}}{\varepsilon_0}=\frac{\mu'_{s}}{\mu_0}= 
	\begin{bmatrix}
	\alpha_{11} &
	\alpha_{12}&
	0 \\
	\alpha_{21} &
	
	\alpha_{22}&
	0\\
	0 &
	0 &
	\alpha_{33}
	\end{bmatrix} 
	\end{equation}
\end{subequations}
where the coefficients of $\alpha_{ij}$ are
\begin{align}
&\alpha_{11} = \frac{(\tau_3 -\tau_2)r^\prime - \tau_3(\tau_1-\tau_2)R(\phi)}{(\tau_3-\tau_2)r^\prime}+ \\ \nonumber
& \frac{\tau^2_3(\tau_1-\tau_2)^2(dR(\phi)/d\phi)^2 }{[(\tau_3-\tau_2)^2 r^\prime - \tau_3 (\tau_1-\tau_2)(\tau_3-\tau_2)R(\phi)]r^\prime} \\ \nonumber
&\alpha_{12} = \alpha_{21}= \frac{(\tau_3(\tau_1-\tau_2)(dR/d\phi))}{(\tau_3-\tau_2)r^\prime- \tau_3 (\tau_1 -\tau_2) R(\phi)}\\ \nonumber
&\alpha_{22} = \frac{(\tau_3-\tau_2)r^\prime}{(\tau_3-\tau_2)r^\prime-\tau_3(\tau_1-\tau_2)R(\phi) } \\ \nonumber
& \alpha_{33}=\frac{(\tau_3-\tau_2)^2 r^\prime -\tau_3(\tau_1-\tau_2)(\tau_3-\tau_2)R(\phi)}{(\tau_3-\tau_1)^2 r^\prime}
\end{align}
As can be seen from Eq.(4b) and Eq.(5), the obtained materials of the stretched region are inhomogeneous and anisotropic with off-diagonal components of $\alpha_{12} / \alpha_{21}$ which cause serious difficulties in their realization process \cite{farhat2016transformation}.
In addition to the inhomogenity that the existence of $R(\phi)$ in the components of $\alpha_{ij}$ dictates, it also demonstrate the dependency of obtained materials to the structure geometry. In other words, an alternation in the coefficients of Eq.(1),which results in new geometry, leads to new materials which must be recalculated. However, since $ R_2(\phi)=\tau_2 \times R(\phi)$  is a fictitious region, $\tau_2$  can achieve any arbitrary value. This will give us a degree of freedom to arbitrarily select the value of $\tau_2$ in a manner that it will eradicate the effect of the off-diagonal components of $\alpha_{12}$ (also $\alpha_{21}$). To this aim, we will set $\tau_2 \rightarrow \tau_3$ as shown in Fig.1.  After performing a quite tedious calculations ( the details are provided in appendix) the new materials for each section will be achieved as
\begin{subequations}
	\begin{equation}
	\label{eq-a}
	\frac{\varepsilon'_{c}}{\varepsilon_0}=\frac{\mu'_{c}}{\mu_0}= 
	\begin{bmatrix}
	1 &
	0&
	0 \\
	0 &
	
	1&
	0\\
	0 &
	0 &
	(\tau_2/ \tau_1)^2
	\end{bmatrix} 
	\end{equation}
	\begin{equation}
	\label{eq-b}
	\frac{\varepsilon'_{s}}{\varepsilon_0}=\frac{\mu'_{s}}{\mu_0}= 
	\begin{bmatrix}
	\infty &
	0&
	0 \\
	0 &
	
	0&
	0\\
	0 &
	0 &
	0
	\end{bmatrix}
	\end{equation}
\end{subequations}
The attained material of Eq.(6b) is known as ONM which were comprehensively discussed earlier. However, for better understanding the physical concept of ONM, the dispersion relation in a medium which is filled with ONM  will be analyzed. In addition, it is noteworthy to mention all the given explanation are for the transverse electric (TE) polarization; however, the same procedure could also be utilized for transverse magnetic (TM) polarization.
It is known that in an anisotropic media, the dispersion relation for transverse electric (TE) polarization incident wave is expressed as \cite{cheng2010radiation}
\begin{equation}
\frac{k^2_r}{\mu_\varphi} + \frac{k^2_\phi}{\mu_r}=k^2_0 \varepsilon_z
\end{equation}
where $k_r$ and $k_\phi$  are the wave vector along the radial and azimuthal directions, respectively and  $k_0$ indicates the wave vector in the free space. 
By utilizing Eq.(6b), it is clear that when  $\mu_r= \infty$, regardless of the value of $k_\phi$, the EM waves will constantly have the wave vector of  $k_r=k_0 \sqrt{\mu_\phi \varepsilon_z}$, that since $\varepsilon_{z}=\mu_{\phi}=0$, the wave vector along radial direction will become zero . Therefore, it is expected that there will be no propagation in the $r$ direction in this kind of material and the EM fields would be mapped point-to-point from the outer boundary to the inner one and vice versa. For further clarifying the concept, several numerical simulations were performed and discussed in the next section.
\section{Results}
 To numerically verify the concept of the propounded method, several arbitrary shape concentrators were simulated for TE polarization, with  \textit{z}-directed E-field, using the COMSOL MULTIPHYSICS finite element solver. Since the outer radiation boundary is a rectangle with the sides of $16\lambda \times 16\lambda$, the provided simulations exhibit both near field and far field behavior of the structure. In addition, for all the performed simulations, the incident wave is assumed to be a plane wave, which is illuminating under the frequency of $ f= 0.38 THz$ and $\tau_i$ are selected to have constant values of $\tau_1 =0.6$ , $\tau_2 =0.99$ and $\tau_3=1$  while the coefficients of  $R(\phi)$ in Eq.(1), will achieve different values for each new case.
 Therefore, by considering these values, the materials of Eq.(6) will be achieved as
 \begin{subequations}
 	\begin{equation}
 	\label{eq-a}
 	\frac{\epsilon'_{c}}{\epsilon_0}=\frac{\mu'_{c}}{\mu_0}= 
 	\begin{bmatrix}
 	1 &
 	0&
 	0 \\
 	0 &
 	
 	1&
 	0\\
 	0 &
 	0 &
 	3.92
 	\end{bmatrix}
 	\end{equation}
 	\begin{equation}
 	\label{eq-b}
 	\frac{\epsilon'_{s}}{\epsilon_0}=\frac{\mu'_{s}}{\mu_0}= 
 	\begin{bmatrix}
 	\infty &
 	0&
 	0 \\
 	0 &
 	
 	0&
 	0\\
 	0 &
 	0 &
 	0
 	\end{bmatrix} 
 	\end{equation}
 \end{subequations}
 However, selecting the exact values of $\infty$ and $0$ for the constitutive components cause some errors in the simulation process. To overcome on the mentioned issue, we assumed that $\varepsilon_s/\varepsilon_0 = \mu_s/\mu_0 = diag [1/\xi, \xi, \xi]$, which $diag$ represents a diagonal matrix and $\xi = 0.001$.  Although the value of $\xi$ is not exactly zero, the mentioned approximation of $\xi$ will cause the wave vector along $r$ direction tends to be zero (i.e., $k_r=k_0 \sqrt{\mu_\phi \varepsilon_z}=k_0 \times \xi \rightarrow 0$). Hence, it is acceptable to suppose that the fields are mapping from outer interfaces to the inner ones point-to-point. \par

For the first example, a well known case of circular cross section concentrator is investigated. To this aim, the coefficients of  $R(\phi)$ are simply taken to be independent of $\phi$, that is $R(\phi)= 2.25\lambda$. Hence, by exploiting the obtained materials of Eq.(8), the near field results will be achieved as shown in Fig.2.
 \begin{figure}[!hb]
 	\centering
 	\includegraphics[width=\linewidth]{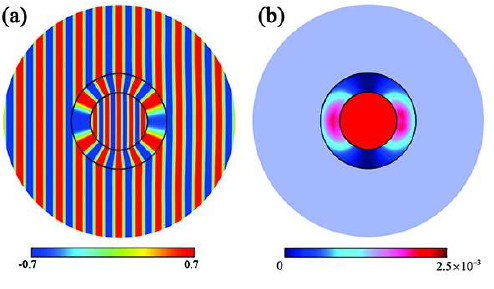}
 	\caption{(a) The electric field distribution of circular cross-section concentrator. (b) Its corresponding power intensity}
 	\label{fig:FIG_3}
 \end{figure}
 \par
 As can be seen from Fig.2(b), in the compressed region the EM intensity is well enhanced. Indeed, it was expected that the power intensity in the mentioned domain would be increased with the ratio of $\tau_2/\tau_1$. This has been verified by the numerical simulation. As it is illustrated, in the free-space where the incident wave is travelling, the power intensity is $1.3 \times 10^{-3} (W/m^2)$ while in the compressed region the intensity is increased as $(\tau_2/\tau_1) \times (1.3 \times 10^{-3})= 2.1 \times 10^{-3} (W/m^2)$, which indicates that the obtained results are well abide with the  ones attained from previously proposed methods\cite{sadeghi2015optical,zhang2011cylindrical}.\par
 However, for the irregular shape concentrators which are important in diverse applications, previously reported approaches have the problems which were mentioned earlier (i.e., inhomogeneous and shape dependent materials). Nevertheless, it will be shown that the attained materials of Eq. (8) are sufficient for any arbitrary value of $R(\phi)$. To this aim, the coefficients of Eq.(1) are varied in a manner that different irregular concentrators are generated. By utilizing the materials of Eq.(8) for each of the new generated shape, the perfect intensity enhancing would be achieved as shown in Fig.3. 

  \begin{figure}[!t]
 	\centering
 	\includegraphics[width=\linewidth]{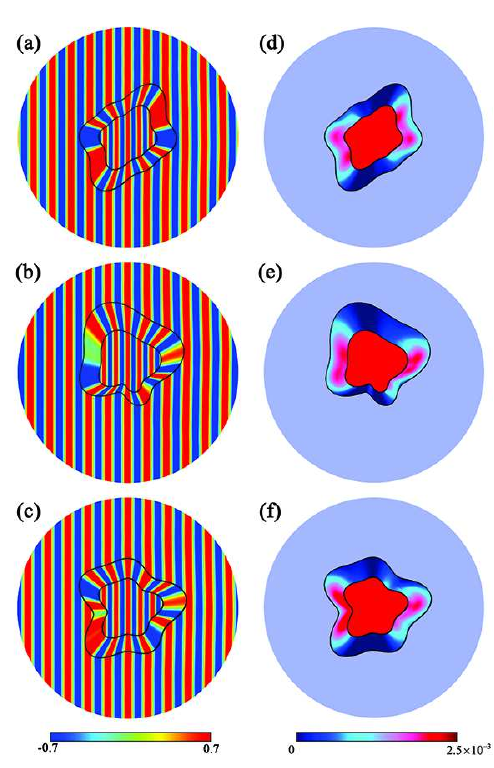}
 	\caption{(a-c) The electric field distribution of different shape concentrators. (d-f) Their corresponding enhanced power intensity}
 	\label{fig:FIG_55}
 \end{figure}
\par
 As it is illustrated in Fig.3, regardless of the input geometry the functionality of the obtained materials remain unchanged. In addition, the same as for the circular shape, the power intensity in the compressed region is increased with the same ratio of $\tau_2 /\tau_1$. However, to have different magnitude in the intensity the ratio of $\tau_2/\tau_1$ must be changed. 
 \vspace{-0.5 cm}
 \section{\label{sec:level5}Conclusion}
 \vspace{-0.5 cm}
 In conclusion, a feasible method for achieving  arbitrary shape EM concentrator that is competent to localize EM waves by obviating the demand of inhomogeneous and anisotropic constitutive parameters is propounded. Several numerical simulations were performed to demonstrate this novel feature of the proposed work.  It was observed that the attained results were well abide with the theoretical predictions. We believe that the newly proposed method in this paper could be utilized in scenario where high concentration of EM waves is of utmost importance. 
 
 \appendix
 \section{Wave equation inside of an optic-null medium}
 As it was mentioned, setting $\tau_2 \rightarrow \tau_3$ will eradicate the effect of off-diagonal components of Eq.(4b). To show this, we will firstly calculate each components\textsc{\char13} value under such an assumption. From Eq. (5) it is clear that when $\tau_2 \rightarrow \tau_3$ each components will be changed to
 \begin{align}
 &\alpha_{11} \rightarrow \infty \\ \nonumber
 &\alpha_{12}=\alpha_{21}= - \frac{dR(\phi)/d\phi}{R(\phi)} \\ \nonumber
 &\alpha_{22} \rightarrow 0 \\ \nonumber
 &\alpha_{33} \rightarrow 0
 \end{align}
 As it is obvious from Eq.(A1), due to the existence of off-diagonal components, the obtained materials under the assumption of $\tau_2 \rightarrow \tau_3$ are still inhomogeneous and anisotropic, which are difficult to be realized. However, according to the Maxwell\textsc{\char13}s equations, it will be demonstrated that the values of $\alpha_{12}$ or $\alpha_{21}$ are not important; thus, they could be considered as zero. To this aim, assume a TE polarized wave is travelling in a medium with the constitutive parameters of
 \begin{subequations}
 	\begin{equation}
 	\label{eq-a}
 	\frac{\varepsilon'}{\varepsilon_0}=
 	\begin{bmatrix}
 	\varepsilon_{rr} &
 	\varepsilon_{r\phi}&
 	0 \\
 	\varepsilon_{r\phi} &
 	
 	\varepsilon_{\phi \phi}&
 	0\\
 	0 &
 	0 &
 	\varepsilon_{zz}
 	\end{bmatrix} 
 	\end{equation}
 	\begin{equation}
 	\label{eq-b}
 	\frac{\mu'}{\mu_0}= 
 	\begin{bmatrix}
 	\mu_{rr} &
 	\mu_{r\phi}&
 	0 \\
 	\mu_{r\phi} &
 	
 	\mu_{\phi \phi}&
 	0\\
 	0 &
 	0 &
 	\mu_{zz}
 	\end{bmatrix} 
 	\end{equation}
 \end{subequations}
 From Maxwell\textsc{\char13}s equations, the magnetic and electric fields are achieved as
 \begin{subequations}
 	\begin{equation}
 	\label{eq-a}
 	H_r = \frac{1}{j\omega(\mu^2_{r\phi}-\mu_{rr}\mu_{\phi \phi})} \times (\mu_{r\phi} \partial_r  E_z + \frac{\mu_{\phi\phi}}{r}\partial_{\phi} E_z)
 	\end{equation}
 	\begin{equation}
 	\label{eq-b}
 	H_{\phi} = \frac{1}{j\omega(\mu_{rr}\mu_{\phi \phi}-\mu^2_{r\phi})} \times (\mu_{rr} \partial_r  E_z + \frac{\mu_{\phi\phi}}{r}\partial_{\phi} E_z)
 	\end{equation}
 	\begin{equation}
 	\label{eq-c}
 	\frac{1}{r}\times \partial_r (rH_\phi)-\frac{1}{r} \partial_\phi H_r =j \omega \varepsilon_{zz} E_z
 	\end{equation}
 \end{subequations}
 Substituting Eq.(A3a) and Eq. (A3b) in Eq. (A3c) the governing equation for electric field in such a medium will be attained as
 \begin{align}
 &\mu_{rr}\partial^2_{r}E_z +  \frac{\mu_{rr}}{r}\partial_{r}E_z+ 2\frac{\mu_{r\phi}}{r}\partial_{r\phi}E_z+ \\ \nonumber
 & \frac{\mu_{\phi\phi}}{r^2}\partial^2_{\phi}E_z+\omega^2 (\mu_{rr}\mu_{\phi \phi}-\mu^2_{r\phi}) \varepsilon_{zz} E_z =0 
 \end{align}
 
 The attained expression of Eq.(A4) is the general form of the electric field in an anisotropic medium.
 Assume $\mu_{rr}=1/\xi$, $\mu_{\phi \phi}= \xi$ and $\varepsilon_{zz}= \xi$, with $\xi \rightarrow 0$.  Substituting these values into Eq.(A4), the governing electric field equation in a medium defined with Eq.(A2) will be attained as
 \begin{align}
 &\frac{1}{\xi}\partial^2_{r}E_z +  \frac{1}{r \xi}\partial_{r}E_z+ 2\frac{\mu_{r\phi}}{r}\partial_{r\phi}E_z+ \\ \nonumber
 & \frac{\xi}{r^2}\partial^2_{\phi}E_z+\omega^2 (1-\mu^2_{r\phi}) \xi E_z =0
 \end{align}
 \vspace{-0.1 cm}
 The above-equation could be easily simplified to a more simplified one as 
 \begin{align}
 &\partial^2_{r}E_z +  \frac{1}{r}\partial_{r}E_z+ 2\frac{\mu_{r\phi} \cdot \xi}{r}\partial_{r\phi}E_z+ \\ \nonumber
 & \frac{\xi^2}{r^2}\partial^2_{\phi}E_z+\omega^2 (1-\mu^2_{r\phi}) \xi^2 E_z =0
 \end{align}
 Since $\xi \rightarrow 0$ and $\mu_{r\phi}$ has finite value of $\frac{dR(\phi)/d\phi}{R(\phi)}$ according to Eq.(A1) (also $\varepsilon_{r\phi} $ for TM polarization); hence, the exact values of $\mu_{r\phi}$ is not important.  This is because only the products of these values play crucial role as shown in the wave equation (i.e., Eq.(A6)) not each of them individually. Therefore, one can assume any desirable finite value for the off-diagonal components. Here we assumed $\varepsilon_{r\phi} = \mu_{r\phi}=0$.  Therefore, the final materials, which describe the performance of an arbitrary shape concentrator, would be achieved as the ones given in Eq.(6).
 
\bibliography{Ref}

\end{document}